\documentclass[
reprint,nofootinbib,amsmath,amssymb,aps,preprintnumbers,
]{revtex4-1}

\usepackage{graphicx}
\usepackage{bm}
\usepackage{braket}
\usepackage{cases} 
\usepackage{here}
\usepackage{color}
\usepackage{upgreek}

\begin{document}

\preprint{KOBE-COSMO-19-01}

\title{Probing GHz Gravitational Waves with Graviton-magnon Resonance}

\author{Asuka Ito}
\email[]{asuka-ito@stu.kobe-u.ac.jp}

\author{Tomonori Ikeda}
\email[]{152s102s@stu.kobe-u.ac.jp}

\author{Kentaro Miuchi}
\email[]{miuchi@phys.sci.kobe-u.ac.jp}

\author{Jiro Soda }
\email[]{jiro@phys.sci.kobe-u.ac.jp}

\affiliation{Department of Physics, Kobe University, Kobe 657-8501, Japan}

\date{\today}

\begin{abstract}
A novel method for extending the frequency frontier in gravitational wave observations is proposed. 
It is shown that gravitational waves can excite a magnon.
Thus, gravitational waves  can be probed by a graviton-magnon detector which measures resonance fluorescence of magnons.
Searching for gravitational waves with a wave length $\lambda$ by using a ferromagnetic sample 
with a dimension $l$,
the sensitivity of the graviton-magnon detector reaches spectral densities, around
$5.4 \times 10^{-22} \times (\frac{l}{\lambda /2\pi})^{-2} \  [{\rm Hz}^{-1/2}]$ at 14 GHz and
$8.6 \times 10^{-21} \times (\frac{l}{\lambda /2\pi})^{-2} \  [{\rm Hz}^{-1/2}]$ at 8.2 GHz, respectively.


\end{abstract}

\maketitle

%
%
%
%
%
%
%
%
\section{Introduction}
In 2015, the gravitational wave interferometer detector LIGO~\cite{Abbott:2016blz} opened up full-blown multi-messenger astronomy and cosmology, where electromagnetic waves, gravitational waves, neutrinos, and cosmic rays are
utilized to explore the universe.
In future, as the history of electromagnetic wave astronomy tells us, 
multi-frequency gravitational wave observations will be required
to boost the multi-messenger astronomy and cosmology.

The purpose of this letter is to present a novel idea for extending the frequency frontier in gravitational wave observations and to report the first limit on GHz gravitational waves.
As we will see below, there are experimental and theoretical motivations to probe GHz gravitational waves. 

It is useful to review the current status of gravitational wave observations~\cite{Kuroda:2015owv}.
It should be stressed that there exists a lowest measurable frequency.
Indeed, the lowest frequency we can measure is around $10^{-18}$ Hz
below which the wave length of gravitational waves 
exceeds the current Hubble horizon.
Measuring the temperature anisotropy and the B-mode polarization of the cosmic microwave background~\cite{Akrami:2018odb,Ade:2015tva}, we can probe gravitational waves with frequencies between $10^{-18}$ Hz and $10^{-16}$ Hz.
Astrometry of extragalactic radio sources is sensitive to gravitational waves with frequencies between $10^{-16}$ Hz and $10^{-9}$ Hz~\cite{Gwinn:1996gv,Darling:2018hmc}. 
The pulsar timing arrays, 
like EPTA~\cite{Lentati:2015qwp,Babak:2015lua},
IPTA~\cite{Perera:2019sca} and
NANOGrav~\cite{Arzoumanian:2018saf},
observe the gravitational waves in the frequency band 
from $10^{-9}$ Hz to $10^{-7}$ Hz.
Doppler tracking of a space craft, which uses
a measurement similar to the pulsar timing arrays, can search for gravitational waves 
in the frequency band from $10^{-7}$ Hz to $10^{-3}$ Hz~\cite{Armstrong:2003ay}.
The space interferometers LISA~\cite{AmaroSeoane:2012km} and DECIGO~\cite{Seto:2001qf} can cover the range
between $10^{-3}$ Hz and $10$ Hz.
The interferometer detectors LIGO~\cite{LIGO}, Virgo~\cite{Virgo}, and KAGRA~\cite{Somiya:2011np}  
with km size arm lengths
can search for gravitational waves with frequencies from $10$ Hz to $1$ kHz. 
In this frequency band, resonant bar experiments~\cite{Maggiore:1999vm} are
complementary to the interferometers~\cite{Acernese:2007ek}.
Furthermore,  interferometers  can be used to measure 
gravitational waves with
the frequencies between $1$ kHz and $100$ MHz.
In fact, recently, the limit on gravitational waves at MHz was reported~\cite{Chou:2016hbb}. To our best knowledge, the measurement of 
$100$ MHz gravitational waves with a $0.75$ m arm length interferometer~\cite{Akutsu:2008qv} 
 is the highest frequency gravitational wave experiment to date.
Thus, the frequency range higher than $100$ MHz  is 
remaining to be explored.
Given this experimental situation, 
GHz experiments are desired to extend the frequency frontier.

Theoretically, GHz gravitational waves are interesting
 from various points of view.
As is well known, inflation can produce primordial gravitational waves. Among the features of
primordial gravitational waves, the most clear signature is the break of the spectrum determined by the energy scale of the inflation,
which may locate at around GHz~\cite{Maggiore:1999vm}.  
Moreover, at the end of inflation or just after inflation, there may be a high frequency peak 
of gravitational waves~\cite{Ito:2016aai,Khlebnikov:1997di}.
Remarkably, there is a chance to observe non-classical nature of
primordial gravitational waves with frequency between MHz and GHz~\cite{Kanno:2018cuk}.
On the other hand, there are many astrophysical sources producing high frequency gravitational waves~\cite{BisnovatyiKogan:2004bk}.
Among them, primordial black holes are the most interesting ones because they give a hint of the 
information loss problem.
Exotic signals from extra dimensions may exist in the GHz band~\cite{Seahra:2004fg,Clarkson:2006pq}.
Hence, GHz gravitational waves could be a window to the extra dimensions~\cite{Ishihara:2007ni}.
Thus, it is worth investigating GHz gravitational waves to understand 
the astrophysical process, the early universe, and quantum gravity.

In this letter, we propose a novel method for detecting GHz gravitational waves with a magnon detector.
First, we show that gravitational waves excite 
magnons in a ferromagnetic insulator.
Furthermore, using experimental results of measurement of resonance fluorescence of 
magnons~\cite{Crescini:2018qrz,Flower:2018qgb},
we demonstrate that the sensitivity to
the spectral density of gravitational waves are around
$5.4 \times 10^{-22} \times (\frac{l}{\lambda /2\pi})^{-2}\  [{\rm Hz}^{-1/2}]$ at 14 GHz and
$8.6 \times 10^{-21} \times (\frac{l}{\lambda /2\pi})^{-2} \  [{\rm Hz}^{-1/2}]$ at 8.2 GHz, respectively,
where $l$ is the dimension of the ferromagnetic insulator and $\lambda$ is the wave length of 
the gravitational waves.
%
%
%
%
%
%
%
%
%
\section{Graviton-magnon resonance}
The Dirac equation in curved spacetime with a metric $g_{\mu\nu}$ is given by
\begin{equation}
  i \gamma^{\hat{\alpha}} e^{\mu}_{\hat{\alpha}} \left( \partial_{\mu} + \Gamma_{\mu} +
                 i e A_{\mu} \right) \psi = m \psi \ ,  \label{dira}
\end{equation}
where $\gamma^{\hat{\alpha}}$,  $e$, $A_{\mu}$ are the gamma matrices, the electronic charge,
 and a vector potential, respectively.  A tetrad  $e^{\mu}_{\hat{\alpha}}$ satisfies 
$ e^{\hat{\alpha}}_{\mu} e^{\hat{\beta}}_{\nu} \eta_{\hat{\alpha}\hat{\beta}} = g_{\mu\nu} $. 
The spin connection is defined by
$\Gamma_{\mu} = \frac{1}{2} e^{\hat{\alpha}}_{\nu} \sigma_{\hat{\alpha}\hat{\beta}} 
        \left( \partial_{\mu} e^{\nu\hat{\beta}} + \Gamma^{\nu}_{\lambda\mu} e^{\lambda\hat{\beta}} \right),$
where
$\sigma_{\hat{\alpha}\hat{\beta}} = \frac{1}{4} [ \gamma_{\hat{\alpha}}, \gamma_{\hat{\beta}} ] $ is 
a generator of the Lorentz group and 
$\Gamma^{\mu}_{\nu\lambda}$ is the Christoffel symbol.

In the non-relativistic limit,
one can obtain interaction terms between a magnetic field and a spin as
\begin{equation}
  \mathcal{H}_{{\rm spin}} \simeq - \mu_{B} \left( 2 \delta_{ij} + h_{ij} \right) 
                                    \hat{S}^{i} B^{j}      \ ,  \label{spispi}
\end{equation}
where $\mu_{B} = |e|/{2m}$,  $\hat{\bm{S}}$, and $\bm{B}$  are the Bohr magneton, the spin of the electron,
and an external magnetic field, respectively. 
$h_{ij} (\ll 1)$ describes ``effective'' gravitational waves.%
\footnote{
The discussion should be developed in the proper detector frame.
Then $h_{ij}$ is given by the Riemann tensor like $R_{ikjl} x^{k} x^{l}$, 
where $x^{k}$ is the spatial coordinate of a Fermi normal coordinate~\cite{Manasse:1963zz}.
Consequently, a suppression factor $(\frac{l}{\lambda /2\pi})^{2}$ appears when
we read off the ``true'' gravitational wave from $h_{ij}$ at the final result.
\label{footnote1}
}
The second term shows that gravitational waves can interact with the spin 
in the presence of  external magnetic fields~\cite{Quach:2016uxd}.%

We now consider a ferromagnetic sample which has $N$ electronic spins.
Such a system is well described by the Heisenberg model:
\begin{eqnarray}
  \mathcal{H}_{g} =   - \mu_{B} B_{z} \sum_{i} \left[  2 \hat{S}^{z}_{(i)} 
      +    h_{zj} \hat{S}^{j}_{(i)}  \right] 
       -  \sum_{i,j} J_{ij} \hat{\bm{S}}_{(i)} \cdot \hat{\bm{S}}_{(j)} \ , \nonumber \\  \label{hei}
\end{eqnarray}
where an external magnetic field $B_{z}$ is applied along the $z$-direction and 
$i$ specifies each of the sites of the spins.
The second term represents the interactions
between spins with coupling constants $J_{ij}$. 

Let us consider planar gravitational waves propagating in the $z$-$x$ plane, namely,
the wave number vector of the gravitational waves $\bm{k}$ has a direction $\hat{k} = ( \sin\theta , 0 , \cos\theta )$.
Moreover,  we assume that the wave length of the gravitational waves is much longer than the dimension of the sample.
This is the case of cavity experiments which we utilize in the next section.
We can expand the metric perturbations in terms of linear polarization tensors satisfying 
$e^{(\sigma)}_{ij} e^{(\sigma')}_{ij} = \delta_{\sigma\sigma'}$ as
\begin{equation}
  h_{ij}(t) = h^{(+)}(t) e^{(+)}_{ij} + h^{(\times)}(t) e^{(\times)}_{ij} \ , \label{expan}
\end{equation}
where we used the fact that the amplitude is approximately uniform over the sample.
More explicitly, we took the representation 
\begin{numcases}
  {}
  h^{(+)}(t) = \frac{h^{(+)}}{2} \left( e^{-iw_{h}t} + e^{iw_{h}t} \right) \ , & \\
  h^{(\times)}(t) = \frac{h^{(\times)}}{2} 
                      \left( e^{-i(w_{h}t + \alpha)} + e^{i(w_{h}t +\alpha)} \right) \ , \label{cross} & 
\end{numcases}
where $\omega_{h}$ is an angular frequency of the gravitational waves and $\alpha$ represents 
a difference of the phases of polarizations.
Note that the polarization tensors  can be explicitly constructed as
\begin{eqnarray}
  e _{ij}^{(+)} &=& \frac{1}{\sqrt{2}}\left(
    \begin{array}{ccc}
      \cos\theta^{2} & 0 & -\cos\theta \sin\theta \\
      0 & -1 & 0 \\
      -\cos\theta \sin\theta & 0 & \sin\theta^{2}
    \end{array} 
  \right) , \label{lipo1}  \\
  e _{ij}^{(\times)} &=& \frac{1}{\sqrt{2}}\left(
    \begin{array}{ccc}
      0 & \cos\theta & 0 \\
      \cos\theta & 0 & -\sin\theta \\
      0 & -\sin\theta & 0
    \end{array} 
  \right) .  \label{lipo2}
\end{eqnarray}
In the above Eqs.\,(\ref{lipo1}) and (\ref{lipo2}), 
we defined the $+$ mode as a deformation in the $y$-direction.

It is well known that the spin system (\ref{hei}) can be rewritten by using
 the Holstein-Primakoff transformation~\cite{Holstein:1940zp}: 
\begin{eqnarray}
  \begin{cases}
   \hat{S}_{(i)}^{z} = \frac{1}{2} - \hat{C}_{i}^{\dagger} \hat{C}_{i} \ , & \\
   \hat{S}_{(i)}^{+} = \sqrt{1- \hat{C}_{i}^{\dagger} \hat{C}_{i}} \ \hat{C}_{i}  \ , & \\
   \hat{S}_{(i)}^{-} = \hat{C}_{i}^{\dagger} \sqrt{1- \hat{C}_{i}^{\dagger} \hat{C}_{i}} \ , &
  \end{cases} \label{pri}
\end{eqnarray}
where the bosonic operators $\hat{C}_{i}$ and $\hat{C}^{\dagger}_{i}$ 
satisfy the commutation relations $[\hat{C}_{i}, \hat{C}^{\dagger}_{j}]=\delta_{ij}$ and
$S_{(j)}^{\pm} = S_{(j)}^{x} \pm i S_{(j)}^{y}$ are the ladder operators.
The bosonic operators describe spin waves with dispersion relations determined by 
$B_{z}$ and $J_{ij}$. 
Furthermore, provided that contributions from the surface of the sample are negligible,
one can expand the bosonic operators by plane waves as 
\begin{equation}
  \hat{C}_{i} = \sum_{\bm{k}} \frac{e^{-i\bm{k} \cdot \bm{r}_{i}}}{\sqrt{N}} \hat{c}_{k} \ , \label{bbb}
\end{equation}
where $\bm{r}_{i}$ is the position vector of the $i$ spin.
The excitation of the spin waves created by $\hat{c}_{k}^{\dagger}$ is called a magnon.
From now on, we only consider the uniform mode of magnons.
Then from Eq.\,(\ref{hei}), using 
the rotating wave approximation and assuming $\hat{c}^{\dagger} \hat{c} \ll 1$,
one can deduce
\begin{equation}
   \mathcal{H}_{g} \simeq 2\mu_{B} B_{z}  \hat{c}^{\dagger} \hat{c} +  
 g_{eff} \left(    \hat{c}^{\dagger} e^{-i\omega_{h}t} +    \hat{c} e^{i\omega_{h}t} \right)  ,
                      \label{dri}
\end{equation}
where $\hat{c} = \hat{c}_{k=0}$ and
\begin{widetext}
\begin{equation}
g_{eff} = \frac{1}{4\sqrt{2}} \mu_{B} B_{z} \sin\theta \sqrt{N} 
\left[ \cos^{2} \theta \, (h^{(+)})^{2} + (h^{(\times)})^{2} + 2 \cos\theta\sin\alpha \, h^{(+)}h^{(\times)} 
\right]^{1/2}  \ , \label{effco}
\end{equation}
is an effective coupling constant between gravitational waves and a magnon.
From Eq.\,(\ref{effco}), we see that the effective coupling constant has gotten a factor $\sqrt{N}$.
One can also express Eq.\,(\ref{effco}) using the Stokes parameters as 
\begin{equation}
g_{eff} = \frac{1}{4\sqrt{2}} \mu_{B} B_{z} \sin\theta \sqrt{N} 
          \left[ \frac{1+\cos^{2}\theta}{2} \, I - 
          \frac{\sin^{2}\theta}{2} \, Q  + \cos \theta \, V  \right]^{1/2}  \ , \label{effco2}
\end{equation}
\end{widetext}
where the Stokes parameters for gravitational waves are defined by 
\begin{eqnarray}
  \begin{cases}
   I = (h^{(+)})^{2} + (h^{(\times)})^{2} \ , & \\
   Q = (h^{(+)})^{2} - (h^{(\times)})^{2}  \ , & \\
   U = 2 \cos\alpha \, h^{(+)} h^{(\times)}  \ , & \\
   V = 2 \sin\alpha \, h^{(+)} h^{(\times)}  \ . &
  \end{cases} \label{stokes}
\end{eqnarray}
They satisfy $I^{2} = U^{2} + Q^{2} +V^{2}$.
We see that the effective coupling constant depends on the polarizations.
Note that the Stokes parameters $Q$ and $U$ transform as 
\begin{equation}
\binom{Q'}{U'} = \begin{pmatrix}
\cos 4\Psi & \sin 4\Psi \\
-\sin 4\Psi & \cos 4\Psi   \end{pmatrix}
\binom{Q}{U}
\end{equation}
where $\Psi$ is the rotation angle around $\bm{k}$.

The second term in Eq.\,(\ref{dri}) shows that 
 planar gravitational waves induce the resonant spin precessions
if the angular frequency of the gravitational waves is near the 
Lamor frequency, $2\mu_{B} B_{z}$.
It is worth noting that the situation is similar to the resonant bar experiments~\cite{Maggiore:1999vm} where 
planar gravitational waves excite phonons in a bar detector.

In the next section, utilizing the graviton-magnon resonance, we will give 
upper limits on GHz gravitational waves.
\section{Limits on GHz gravitational waves}
In the previous section, we showed that planar gravitational waves can induce resonant spin precession of electrons.
It is our observation  that the same resonance is caused by coherent oscillation of the axion dark 
matter~\cite{Barbieri:1985cp}.
Recently, measurements of resonance fluorescence of magnons induced by the axion dark matter 
was conducted and upper bounds on an axion-electron coupling constant have been 
obtained~\cite{Crescini:2018qrz,Flower:2018qgb}.
The point is that we can utilize these experimental results to give the upper bounds on the amplitude of GHz gravitational waves.

Actually, the interaction hamiltonian which describe an axion-magnon resonance is given by
\begin{equation}
  \mathcal{H}_{a} 
  = \tilde{g}_{eff} \left(    \hat{c}^{\dagger} e^{-im_{a}t} +    \hat{c} e^{im_{a}t} \right) \ , \label{axion}
\end{equation}
where $\tilde{g}_{eff}$ is an effective coupling constant between an axion and a magnon.
Notice that the axion oscillates with a frequency determined by the axion mass $m_{a}$.
One can see that this form is the same as the interaction term in Eq.\,(\ref{dri}).
Through the hamiltonian (\ref{axion}), 
$\tilde{g}_{eff}$ is related to an axion-electron coupling constant in~\cite{Crescini:2018qrz,Flower:2018qgb}.
Then the axion-electron coupling constant can be converted to 
$\tilde{g}_{eff}$ by using parameters, such as the energy density of the axion dark matter, which 
are explicitly given in~\cite{Crescini:2018qrz,Flower:2018qgb}.
Therefore constraints on $\tilde{g}_{eff}$ (95\% C.L.) can be read from the 
constraints on the axion-electron coupling constant given in~\cite{Crescini:2018qrz} and \cite{Flower:2018qgb},
respectively, as follows:
\begin{eqnarray}
  \tilde{g}_{eff} < 
  \begin{cases}
    3.5 \times 10^{-12} \ {\rm eV}  \ , & \\
    3.1 \times 10^{-11} \ {\rm eV}  \ . &  
  \end{cases} \label{aco}
\end{eqnarray}

\begin{figure}[h]
\centering
\includegraphics[width=8.5cm]{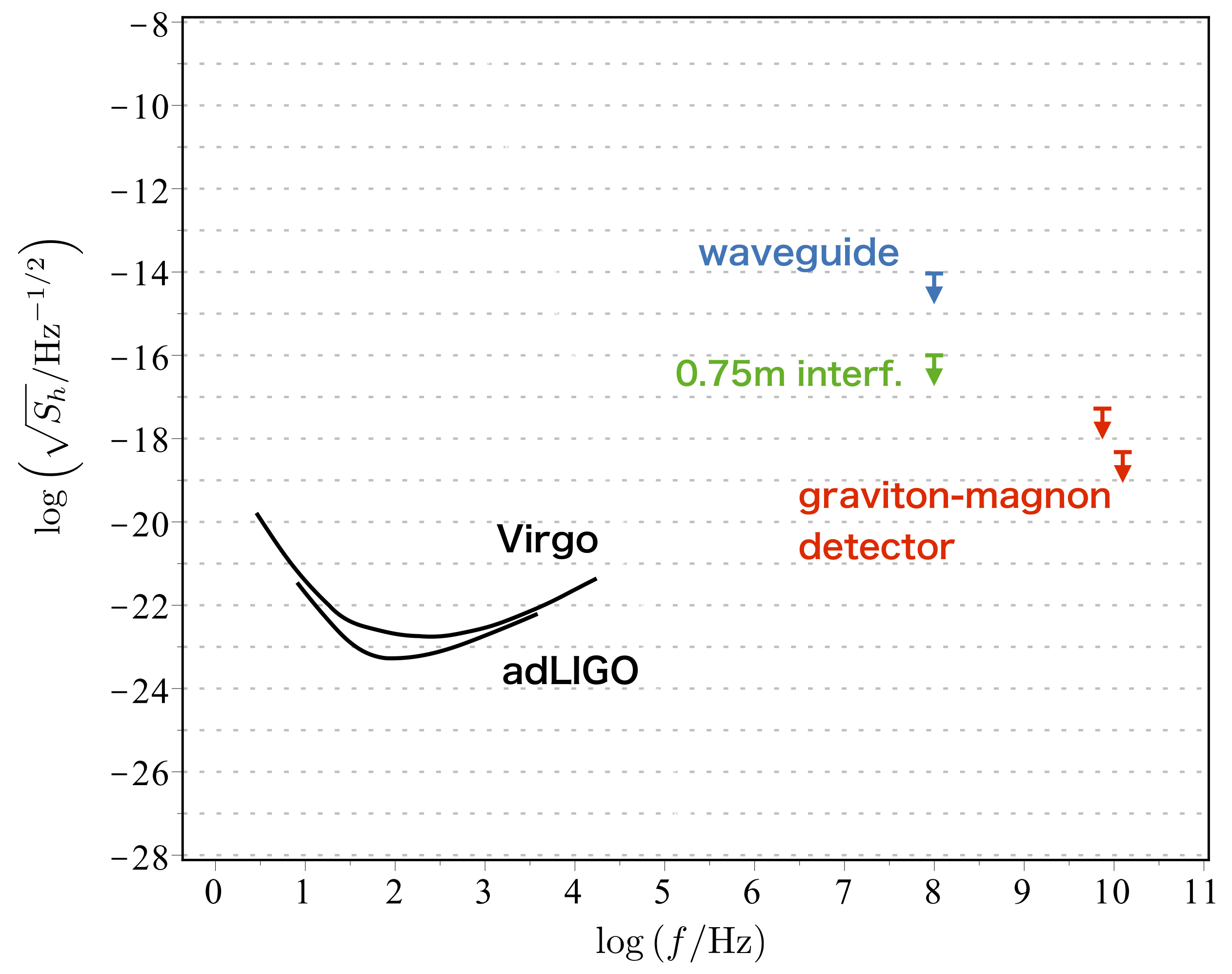}
\caption{Several experimental sensitivities and constraints on high frequency gravitational waves are depicted.
LIGO and Virgo have the sensitivity around $10^{2}$ Hz~\cite{LIGO,Virgo}. 
The blue color represents an upper limit on stochastic gravitational waves
by waveguide experiment using an interaction between 
electromagnetic fields and gravitational waves~\cite{0264-9381-23-22-007}.
The green one is the upper limit on stochastic gravitational waves,
obtained by the 0.75 m interferometer~\cite{Akutsu:2008qv}.
The red color represents the sensitivity of the graviton-magnon detector.}
\label{GWfig}
\end{figure}

It is easy to convert the above constraints 
to those on the amplitude of gravitational waves appearing in the effective coupling constant (\ref{effco2}).
Indeed,  we can read off the external magnetic field $B_{z}$ and the number of electrons $N$ 
as $(B_{z},N) = (0.5 \,{\rm T}, \ 5.6\times 10^{19} )$ from~\cite{Crescini:2018qrz} and 
$(B_{z},N) = (0.3 \,{\rm T}, \ 9.2\times 10^{19} )$ from~\cite{Flower:2018qgb}, respectively.
The external magnetic field $B_{z}$ determines the frequency of gravitational waves we can detect.
Therefore, using Eqs.\,(\ref{effco2}), (\ref{aco}) and the above parameters, one can put upper limits on  
gravitational waves at frequencies determined by $B_{z}$.
Since \cite{Crescini:2018qrz} and \cite{Flower:2018qgb} focused on the direction of Cygnus and set the 
external magnetic fields to be perpendicular to it, 
we probe continuous gravitational waves coming from Cygnus with $\theta = \frac{\pi}{2}$
(more precisely, $\sin\theta=0.9$ in \cite{Flower:2018qgb}).
We also assume there to be no linear and circular polarizations, i.e.,  $Q'=U'=V=0$.
Consequently, experimental data \cite{Crescini:2018qrz} and \cite{Flower:2018qgb} show
the sensitivity to the characteristic amplitude of gravitational waves defined by 
$h_{c} = h^{(+)} = h^{(\times)}$ as
\begin{eqnarray}
  h_{c} \sim
  \begin{cases}
    9.1 \times 10^{-17} \times (\frac{l}{\lambda /2\pi})^{-2} \quad {\rm at} \   14  \  {\rm GHz} \ , & \\
    1.1 \times 10^{-15} \times (\frac{l}{\lambda /2\pi})^{-2} \quad {\rm at} \   8.2 \  {\rm GHz}  \ , &
  \end{cases} 
\end{eqnarray}
respectively.
Note that a suppression factor 
$(\frac{l}{\lambda /2\pi})^{2} \sim 10^{-3}$ has appeared (see the footnote \ref{footnote1}).
In terms of the spectral density defined by $S_{h} = h_{c}^{2}/2f$ and 
the energy density parameter defined by $\Omega_{GW} = 2\pi^{2} f^{2} h_{c}^{2} / 3 H_{0}^{2}$ ($H_{0}$ is 
the Hubble parameter),
the sensitivities are
\begin{eqnarray}
  \sqrt{S_{h}} \sim
  \begin{cases}
    5.4 \times 10^{-22} \times 
    (\frac{l}{\lambda /2\pi})^{-2} \ [{\rm Hz}^{-1/2}]  \quad {\rm at} \   14  \  {\rm GHz} \ , & \\
    8.6 \times 10^{-21} \times 
    (\frac{l}{\lambda /2\pi})^{-2} \ [{\rm Hz}^{-1/2}]  \quad {\rm at} \   8.2 \  {\rm GHz}  \ . &
  \end{cases}   \label{Cons}
\end{eqnarray}
and 
\begin{eqnarray}
  h_{0}^{2} \Omega_{GW} \sim
  \begin{cases}
    1.1 \times 10^{23} \times (\frac{l}{\lambda /2\pi})^{-2}  \quad {\rm at} \   14  \  {\rm GHz} \ , & \\
    5.3 \times 10^{24} \times (\frac{l}{\lambda /2\pi})^{-2}  \quad {\rm at} \   8.2 \  {\rm GHz}  \ . &
  \end{cases}   \label{aaaaa}
\end{eqnarray}
We depicted the sensitivity on the spectral density with several other gravitational wave experiments 
in Fig.\,\ref{GWfig}.
\section{Discussion}
In this letter, we focused on continuous gravitational waves as an explicit demonstration to show 
the sensitivity of our new gravitational wave detection method as summarized in Fig.\,\ref{GWfig}.
Interestingly, there are several theoretical models predicting high frequency gravitational waves
which are within the scope of our method~\cite{Kuroda:2015owv}.
The graviton-magnon resonance is also useful for probing stochastic gravitational waves with
almost the same sensitivity illustrated in Fig.\,\ref{GWfig}.
Although the current  sensitivity is still not sufficient for putting a meaningful constraint on 
stochastic gravitational waves, 
it is important to pursue the high frequency stochastic gravitational wave search
for future gravitational wave physics.
Moreover, we can probe burst gravitational waves of any wave form
if the duration time is smaller than the relaxation time of a system.
The situation is the same as for resonant bar detectors~\cite{Maggiore:1900zz,Astone:2010mr}.
For instance, in the measurements~\cite{Crescini:2018qrz,Flower:2018qgb}, 
the relaxation time is about $0.1$ $\upmu$s which is
determined by the line width of the ferromagnetic sample and the cavity.
If the duration of a burst of gravitational waves is smaller than $0.1$ $\upmu$s, 
we can detect it.
Furthermore, improving the line width of the sample and the cavity not only leads to 
detecting burst gravitational waves but also to increasing the sensitivity.
As another way to improve sensitivity,
quantum nondemolition measurement may be promising~\cite{Tabuchi405,TABUCHI2016729,Lachance-Quirione1603150}.
In particular, although we assumed that a gravitational wave was approximately monochromatic, 
there might be cases where the approximation is not valid.
In such cases, quantum nondemolition measurement would be useful.
\section{Conclusion}
Given the importance of extending the frequency frontier in gravitational wave observations, 
we proposed a novel method to detect GHz gravitational waves with  the magnon detector.
Indeed, gravitational waves can excite a magnon.
Using experimental results for the axion dark matter search, \cite{Crescini:2018qrz} and \cite{Flower:2018qgb},
we showed that the sensitivity to the spectral density of continuous gravitational waves 
reaches around
$5.4 \times 10^{-22} \times (\frac{l}{\lambda /2\pi})^{-2} \  [{\rm Hz}^{-1/2}]$ at 14 GHz and
$8.6 \times 10^{-21} \times (\frac{l}{\lambda /2\pi})^{-2} \  [{\rm Hz}^{-1/2}]$ at 8.2 GHz, respectively.
One can perform all sky search of continuous gravitational waves at the above sensitivity with the graviton-magnon detector.
We can also search for stochastic gravitational waves
and burst gravitational waves with almost the same sensitivity.
\begin{acknowledgments}
A.\,I. was supported by Grant-in-Aid for JSPS Research Fellow and JSPS KAKENHI Grant No.JP17J00216.
J.\,S. was in part supported by JSPS KAKENHI
Grant Numbers JP17H02894, JP17K18778, JP15H05895, JP17H06359, JP18H04589.
J.\,S is also supported by JSPS Bilateral Joint Research
Projects (JSPS-NRF collaboration) “String Axion Cosmology.”
\end{acknowledgments}
\bibliography{GW}

\begin{thebibliography}{39}%
\makeatletter
\providecommand \@ifxundefined [1]{%
 \@ifx{#1\undefined}
}%
\providecommand \@ifnum [1]{%
 \ifnum #1\expandafter \@firstoftwo
 \else \expandafter \@secondoftwo
 \fi
}%
\providecommand \@ifx [1]{%
 \ifx #1\expandafter \@firstoftwo
 \else \expandafter \@secondoftwo
 \fi
}%
\providecommand \natexlab [1]{#1}%
\providecommand \enquote  [1]{``#1''}%
\providecommand \bibnamefont  [1]{#1}%
\providecommand \bibfnamefont [1]{#1}%
\providecommand \citenamefont [1]{#1}%
\providecommand \href@noop [0]{\@secondoftwo}%
\providecommand \href [0]{\begingroup \@sanitize@url \@href}%
\providecommand \@href[1]{\@@startlink{#1}\@@href}%
\providecommand \@@href[1]{\endgroup#1\@@endlink}%
\providecommand \@sanitize@url [0]{\catcode `\\12\catcode `\$12\catcode
  `\&12\catcode `\#12\catcode `\^12\catcode `\_12\catcode `\%12\relax}%
\providecommand \@@startlink[1]{}%
\providecommand \@@endlink[0]{}%
\providecommand \url  [0]{\begingroup\@sanitize@url \@url }%
\providecommand \@url [1]{\endgroup\@href {#1}{\urlprefix }}%
\providecommand \urlprefix  [0]{URL }%
\providecommand \Eprint [0]{\href }%
\providecommand \doibase [0]{http://dx.doi.org/}%
\providecommand \selectlanguage [0]{\@gobble}%
\providecommand \bibinfo  [0]{\@secondoftwo}%
\providecommand \bibfield  [0]{\@secondoftwo}%
\providecommand \translation [1]{[#1]}%
\providecommand \BibitemOpen [0]{}%
\providecommand \bibitemStop [0]{}%
\providecommand \bibitemNoStop [0]{.\EOS\space}%
\providecommand \EOS [0]{\spacefactor3000\relax}%
\providecommand \BibitemShut  [1]{\csname bibitem#1\endcsname}%
\let\auto@bib@innerbib\@empty
\bibitem [{\citenamefont {Abbott}\ \emph {et~al.}(2016)\citenamefont {Abbott}
  \emph {et~al.}}]{Abbott:2016blz}%
  \BibitemOpen
  \bibfield  {author} {\bibinfo {author} {\bibfnamefont {B.~P.}\ \bibnamefont
  {Abbott}} \emph {et~al.} (\bibinfo {collaboration} {Virgo, LIGO
  Scientific}),\ }\href@noop {} {\bibfield  {journal} {\bibinfo  {journal}
  {Phys. Rev. Lett.}\ }\textbf {\bibinfo {volume} {116}},\ \bibinfo {pages}
  {061102} (\bibinfo {year} {2016})},\ \Eprint
  {http://arxiv.org/abs/1602.03837} {arXiv:1602.03837 [gr-qc]} \BibitemShut
  {NoStop}%
\bibitem [{\citenamefont {Kuroda}\ \emph {et~al.}(2015)\citenamefont {Kuroda},
  \citenamefont {Ni},\ and\ \citenamefont {Pan}}]{Kuroda:2015owv}%
  \BibitemOpen
  \bibfield  {author} {\bibinfo {author} {\bibfnamefont {K.}~\bibnamefont
  {Kuroda}}, \bibinfo {author} {\bibfnamefont {W.-T.}\ \bibnamefont {Ni}}, \
  and\ \bibinfo {author} {\bibfnamefont {W.-P.}\ \bibnamefont {Pan}},\
  }\href@noop {} {\bibfield  {journal} {\bibinfo  {journal} {Int. J. Mod.
  Phys.}\ }\textbf {\bibinfo {volume} {D24}},\ \bibinfo {pages} {1530031}
  (\bibinfo {year} {2015})},\ \Eprint {http://arxiv.org/abs/1511.00231}
  {arXiv:1511.00231 [gr-qc]} \BibitemShut {NoStop}%
\bibitem [{\citenamefont {Akrami}\ \emph {et~al.}(2018)\citenamefont {Akrami}
  \emph {et~al.}}]{Akrami:2018odb}%
  \BibitemOpen
  \bibfield  {author} {\bibinfo {author} {\bibfnamefont {Y.}~\bibnamefont
  {Akrami}} \emph {et~al.} (\bibinfo {collaboration} {Planck}),\ }\href@noop {}
  {\  (\bibinfo {year} {2018})},\ \Eprint {http://arxiv.org/abs/1807.06211}
  {arXiv:1807.06211 [astro-ph.CO]} \BibitemShut {NoStop}%
\bibitem [{\citenamefont {Ade}\ \emph {et~al.}(2015)\citenamefont {Ade} \emph
  {et~al.}}]{Ade:2015tva}%
  \BibitemOpen
  \bibfield  {author} {\bibinfo {author} {\bibfnamefont {P.~A.~R.}\
  \bibnamefont {Ade}} \emph {et~al.} (\bibinfo {collaboration} {BICEP2,
  Planck}),\ }\href@noop {} {\bibfield  {journal} {\bibinfo  {journal} {Phys.
  Rev. Lett.}\ }\textbf {\bibinfo {volume} {114}},\ \bibinfo {pages} {101301}
  (\bibinfo {year} {2015})},\ \Eprint {http://arxiv.org/abs/1502.00612}
  {arXiv:1502.00612 [astro-ph.CO]} \BibitemShut {NoStop}%
\bibitem [{\citenamefont {Gwinn}\ \emph {et~al.}(1997)\citenamefont {Gwinn},
  \citenamefont {Eubanks}, \citenamefont {Pyne}, \citenamefont {Birkinshaw},\
  and\ \citenamefont {Matsakis}}]{Gwinn:1996gv}%
  \BibitemOpen
  \bibfield  {author} {\bibinfo {author} {\bibfnamefont {C.~R.}\ \bibnamefont
  {Gwinn}}, \bibinfo {author} {\bibfnamefont {T.~M.}\ \bibnamefont {Eubanks}},
  \bibinfo {author} {\bibfnamefont {T.}~\bibnamefont {Pyne}}, \bibinfo {author}
  {\bibfnamefont {M.}~\bibnamefont {Birkinshaw}}, \ and\ \bibinfo {author}
  {\bibfnamefont {D.~N.}\ \bibnamefont {Matsakis}},\ }\href@noop {} {\bibfield
  {journal} {\bibinfo  {journal} {Astrophys. J.}\ }\textbf {\bibinfo {volume}
  {485}},\ \bibinfo {pages} {87} (\bibinfo {year} {1997})},\ \Eprint
  {http://arxiv.org/abs/astro-ph/9610086} {arXiv:astro-ph/9610086 [astro-ph]}
  \BibitemShut {NoStop}%
\bibitem [{\citenamefont {Darling}\ \emph {et~al.}(2018)\citenamefont
  {Darling}, \citenamefont {Truebenbach},\ and\ \citenamefont
  {Paine}}]{Darling:2018hmc}%
  \BibitemOpen
  \bibfield  {author} {\bibinfo {author} {\bibfnamefont {J.}~\bibnamefont
  {Darling}}, \bibinfo {author} {\bibfnamefont {A.~E.}\ \bibnamefont
  {Truebenbach}}, \ and\ \bibinfo {author} {\bibfnamefont {J.}~\bibnamefont
  {Paine}},\ }\href@noop {} {\bibfield  {journal} {\bibinfo  {journal}
  {Astrophys. J.}\ }\textbf {\bibinfo {volume} {861}},\ \bibinfo {pages} {113}
  (\bibinfo {year} {2018})},\ \Eprint {http://arxiv.org/abs/1804.06986}
  {arXiv:1804.06986 [astro-ph.IM]} \BibitemShut {NoStop}%
\bibitem [{\citenamefont {Lentati}\ \emph {et~al.}(2015)\citenamefont {Lentati}
  \emph {et~al.}}]{Lentati:2015qwp}%
  \BibitemOpen
  \bibfield  {author} {\bibinfo {author} {\bibfnamefont {L.}~\bibnamefont
  {Lentati}} \emph {et~al.},\ }\href@noop {} {\bibfield  {journal} {\bibinfo
  {journal} {Mon. Not. Roy. Astron. Soc.}\ }\textbf {\bibinfo {volume} {453}},\
  \bibinfo {pages} {2576} (\bibinfo {year} {2015})},\ \Eprint
  {http://arxiv.org/abs/1504.03692} {arXiv:1504.03692 [astro-ph.CO]}
  \BibitemShut {NoStop}%
\bibitem [{\citenamefont {Babak}\ \emph {et~al.}(2016)\citenamefont {Babak}
  \emph {et~al.}}]{Babak:2015lua}%
  \BibitemOpen
  \bibfield  {author} {\bibinfo {author} {\bibfnamefont {S.}~\bibnamefont
  {Babak}} \emph {et~al.},\ }\href@noop {} {\bibfield  {journal} {\bibinfo
  {journal} {Mon. Not. Roy. Astron. Soc.}\ }\textbf {\bibinfo {volume} {455}},\
  \bibinfo {pages} {1665} (\bibinfo {year} {2016})},\ \Eprint
  {http://arxiv.org/abs/1509.02165} {arXiv:1509.02165 [astro-ph.CO]}
  \BibitemShut {NoStop}%
\bibitem [{\citenamefont {Perera}\ \emph {et~al.}(2019)\citenamefont {Perera}
  \emph {et~al.}}]{Perera:2019sca}%
  \BibitemOpen
  \bibfield  {author} {\bibinfo {author} {\bibfnamefont {B.~B.~P.}\
  \bibnamefont {Perera}} \emph {et~al.},\ }\href {\doibase
  10.1093/mnras/stz2857} {\bibfield  {journal} {\bibinfo  {journal} {Mon. Not.
  Roy. Astron. Soc.}\ }\textbf {\bibinfo {volume} {490}},\ \bibinfo {pages}
  {4666} (\bibinfo {year} {2019})},\ \Eprint {http://arxiv.org/abs/1909.04534}
  {arXiv:1909.04534 [astro-ph.HE]} \BibitemShut {NoStop}%
\bibitem [{\citenamefont {Arzoumanian}\ \emph {et~al.}(2018)\citenamefont
  {Arzoumanian} \emph {et~al.}}]{Arzoumanian:2018saf}%
  \BibitemOpen
  \bibfield  {author} {\bibinfo {author} {\bibfnamefont {Z.}~\bibnamefont
  {Arzoumanian}} \emph {et~al.} (\bibinfo {collaboration} {NANOGRAV}),\
  }\href@noop {} {\bibfield  {journal} {\bibinfo  {journal} {Astrophys. J.}\
  }\textbf {\bibinfo {volume} {859}},\ \bibinfo {pages} {47} (\bibinfo {year}
  {2018})},\ \Eprint {http://arxiv.org/abs/1801.02617} {arXiv:1801.02617
  [astro-ph.HE]} \BibitemShut {NoStop}%
\bibitem [{\citenamefont {Armstrong}\ \emph {et~al.}(2003)\citenamefont
  {Armstrong}, \citenamefont {Iess}, \citenamefont {Tortora},\ and\
  \citenamefont {Bertotti}}]{Armstrong:2003ay}%
  \BibitemOpen
  \bibfield  {author} {\bibinfo {author} {\bibfnamefont {J.~W.}\ \bibnamefont
  {Armstrong}}, \bibinfo {author} {\bibfnamefont {L.}~\bibnamefont {Iess}},
  \bibinfo {author} {\bibfnamefont {P.}~\bibnamefont {Tortora}}, \ and\
  \bibinfo {author} {\bibfnamefont {B.}~\bibnamefont {Bertotti}},\ }\href
  {\doibase 10.1086/379505} {\bibfield  {journal} {\bibinfo  {journal}
  {Astrophys. J.}\ }\textbf {\bibinfo {volume} {599}},\ \bibinfo {pages} {806}
  (\bibinfo {year} {2003})}\BibitemShut {NoStop}%
\bibitem [{\citenamefont {Amaro-Seoane}\ \emph {et~al.}(2013)\citenamefont
  {Amaro-Seoane} \emph {et~al.}}]{AmaroSeoane:2012km}%
  \BibitemOpen
  \bibfield  {author} {\bibinfo {author} {\bibfnamefont {P.}~\bibnamefont
  {Amaro-Seoane}} \emph {et~al.},\ }\href@noop {} {\bibfield  {journal}
  {\bibinfo  {journal} {GW Notes}\ }\textbf {\bibinfo {volume} {6}},\ \bibinfo
  {pages} {4} (\bibinfo {year} {2013})},\ \Eprint
  {http://arxiv.org/abs/1201.3621} {arXiv:1201.3621 [astro-ph.CO]} \BibitemShut
  {NoStop}%
\bibitem [{\citenamefont {Seto}\ \emph {et~al.}(2001)\citenamefont {Seto},
  \citenamefont {Kawamura},\ and\ \citenamefont {Nakamura}}]{Seto:2001qf}%
  \BibitemOpen
  \bibfield  {author} {\bibinfo {author} {\bibfnamefont {N.}~\bibnamefont
  {Seto}}, \bibinfo {author} {\bibfnamefont {S.}~\bibnamefont {Kawamura}}, \
  and\ \bibinfo {author} {\bibfnamefont {T.}~\bibnamefont {Nakamura}},\ }\href
  {\doibase 10.1103/PhysRevLett.87.221103} {\bibfield  {journal} {\bibinfo
  {journal} {Phys. Rev. Lett.}\ }\textbf {\bibinfo {volume} {87}},\ \bibinfo
  {pages} {221103} (\bibinfo {year} {2001})},\ \Eprint
  {http://arxiv.org/abs/astro-ph/0108011} {arXiv:astro-ph/0108011 [astro-ph]}
  \BibitemShut {NoStop}%
\bibitem [{LIG()}]{LIGO}%
  \BibitemOpen
  \href@noop {} {\enquote {\bibinfo {title} {Ligo},}\ }\bibinfo {howpublished}
  {https://www.ligo.caltech.edu/page/study-work}\BibitemShut {NoStop}%
\bibitem [{Vir()}]{Virgo}%
  \BibitemOpen
  \href@noop {} {\enquote {\bibinfo {title} {Virgo},}\ }\bibinfo {howpublished}
  {http://www.virgo-gw.eu/}\BibitemShut {NoStop}%
\bibitem [{\citenamefont {Somiya}(2012)}]{Somiya:2011np}%
  \BibitemOpen
  \bibfield  {author} {\bibinfo {author} {\bibfnamefont {K.}~\bibnamefont
  {Somiya}} (\bibinfo {collaboration} {KAGRA}),\ }\href {\doibase
  10.1088/0264-9381/29/12/124007} {\bibfield  {journal} {\bibinfo  {journal}
  {Class. Quant. Grav.}\ }\textbf {\bibinfo {volume} {29}},\ \bibinfo {pages}
  {124007} (\bibinfo {year} {2012})},\ \Eprint {http://arxiv.org/abs/1111.7185}
  {arXiv:1111.7185 [gr-qc]} \BibitemShut {NoStop}%
\bibitem [{\citenamefont {Maggiore}(2000)}]{Maggiore:1999vm}%
  \BibitemOpen
  \bibfield  {author} {\bibinfo {author} {\bibfnamefont {M.}~\bibnamefont
  {Maggiore}},\ }\href@noop {} {\bibfield  {journal} {\bibinfo  {journal}
  {Phys. Rept.}\ }\textbf {\bibinfo {volume} {331}},\ \bibinfo {pages} {283}
  (\bibinfo {year} {2000})},\ \Eprint {http://arxiv.org/abs/gr-qc/9909001}
  {arXiv:gr-qc/9909001 [gr-qc]} \BibitemShut {NoStop}%
\bibitem [{\citenamefont {Acernese}\ \emph {et~al.}(2008)\citenamefont
  {Acernese} \emph {et~al.}}]{Acernese:2007ek}%
  \BibitemOpen
  \bibfield  {author} {\bibinfo {author} {\bibfnamefont {F.}~\bibnamefont
  {Acernese}} \emph {et~al.} (\bibinfo {collaboration} {VIRGO
  AURIGA-EXPLORER-NAUTILUS}),\ }\href@noop {} {\bibfield  {journal} {\bibinfo
  {journal} {Class. Quant. Grav.}\ }\textbf {\bibinfo {volume} {25}},\ \bibinfo
  {pages} {205007} (\bibinfo {year} {2008})},\ \Eprint
  {http://arxiv.org/abs/0710.3752} {arXiv:0710.3752 [gr-qc]} \BibitemShut
  {NoStop}%
\bibitem [{\citenamefont {Chou}\ \emph {et~al.}(2017)\citenamefont {Chou} \emph
  {et~al.}}]{Chou:2016hbb}%
  \BibitemOpen
  \bibfield  {author} {\bibinfo {author} {\bibfnamefont {A.~S.}\ \bibnamefont
  {Chou}} \emph {et~al.} (\bibinfo {collaboration} {Holometer}),\ }\href@noop
  {} {\bibfield  {journal} {\bibinfo  {journal} {Phys. Rev.}\ }\textbf
  {\bibinfo {volume} {D95}},\ \bibinfo {pages} {063002} (\bibinfo {year}
  {2017})},\ \Eprint {http://arxiv.org/abs/1611.05560} {arXiv:1611.05560
  [astro-ph.IM]} \BibitemShut {NoStop}%
\bibitem [{\citenamefont {Akutsu}\ \emph {et~al.}(2008)\citenamefont {Akutsu}
  \emph {et~al.}}]{Akutsu:2008qv}%
  \BibitemOpen
  \bibfield  {author} {\bibinfo {author} {\bibfnamefont {T.}~\bibnamefont
  {Akutsu}} \emph {et~al.},\ }\href@noop {} {\bibfield  {journal} {\bibinfo
  {journal} {Phys. Rev. Lett.}\ }\textbf {\bibinfo {volume} {101}},\ \bibinfo
  {pages} {101101} (\bibinfo {year} {2008})},\ \Eprint
  {http://arxiv.org/abs/0803.4094} {arXiv:0803.4094 [gr-qc]} \BibitemShut
  {NoStop}%
\bibitem [{\citenamefont {Ito}\ and\ \citenamefont {Soda}(2016)}]{Ito:2016aai}%
  \BibitemOpen
  \bibfield  {author} {\bibinfo {author} {\bibfnamefont {A.}~\bibnamefont
  {Ito}}\ and\ \bibinfo {author} {\bibfnamefont {J.}~\bibnamefont {Soda}},\
  }\href {\doibase 10.1088/1475-7516/2016/04/035} {\bibfield  {journal}
  {\bibinfo  {journal} {JCAP}\ }\textbf {\bibinfo {volume} {1604}},\ \bibinfo
  {pages} {035} (\bibinfo {year} {2016})},\ \Eprint
  {http://arxiv.org/abs/1603.00602} {arXiv:1603.00602 [hep-th]} \BibitemShut
  {NoStop}%
\bibitem [{\citenamefont {Khlebnikov}\ and\ \citenamefont
  {Tkachev}(1997)}]{Khlebnikov:1997di}%
  \BibitemOpen
  \bibfield  {author} {\bibinfo {author} {\bibfnamefont {S.~Y.}\ \bibnamefont
  {Khlebnikov}}\ and\ \bibinfo {author} {\bibfnamefont {I.~I.}\ \bibnamefont
  {Tkachev}},\ }\href {\doibase 10.1103/PhysRevD.56.653} {\bibfield  {journal}
  {\bibinfo  {journal} {Phys. Rev.}\ }\textbf {\bibinfo {volume} {D56}},\
  \bibinfo {pages} {653} (\bibinfo {year} {1997})},\ \Eprint
  {http://arxiv.org/abs/hep-ph/9701423} {arXiv:hep-ph/9701423 [hep-ph]}
  \BibitemShut {NoStop}%
\bibitem [{\citenamefont {Kanno}\ and\ \citenamefont
  {Soda}(2018)}]{Kanno:2018cuk}%
  \BibitemOpen
  \bibfield  {author} {\bibinfo {author} {\bibfnamefont {S.}~\bibnamefont
  {Kanno}}\ and\ \bibinfo {author} {\bibfnamefont {J.}~\bibnamefont {Soda}},\
  }\href@noop {} {\  (\bibinfo {year} {2018})},\ \Eprint
  {http://arxiv.org/abs/1810.07604} {arXiv:1810.07604 [hep-th]} \BibitemShut
  {NoStop}%
\bibitem [{\citenamefont {Bisnovatyi-Kogan}\ and\ \citenamefont
  {Rudenko}(2004)}]{BisnovatyiKogan:2004bk}%
  \BibitemOpen
  \bibfield  {author} {\bibinfo {author} {\bibfnamefont {G.~S.}\ \bibnamefont
  {Bisnovatyi-Kogan}}\ and\ \bibinfo {author} {\bibfnamefont {V.~N.}\
  \bibnamefont {Rudenko}},\ }\href@noop {} {\bibfield  {journal} {\bibinfo
  {journal} {Class. Quant. Grav.}\ }\textbf {\bibinfo {volume} {21}},\ \bibinfo
  {pages} {3347} (\bibinfo {year} {2004})},\ \Eprint
  {http://arxiv.org/abs/gr-qc/0406089} {arXiv:gr-qc/0406089 [gr-qc]}
  \BibitemShut {NoStop}%
\bibitem [{\citenamefont {Seahra}\ \emph {et~al.}(2005)\citenamefont {Seahra},
  \citenamefont {Clarkson},\ and\ \citenamefont {Maartens}}]{Seahra:2004fg}%
  \BibitemOpen
  \bibfield  {author} {\bibinfo {author} {\bibfnamefont {S.~S.}\ \bibnamefont
  {Seahra}}, \bibinfo {author} {\bibfnamefont {C.}~\bibnamefont {Clarkson}}, \
  and\ \bibinfo {author} {\bibfnamefont {R.}~\bibnamefont {Maartens}},\
  }\href@noop {} {\bibfield  {journal} {\bibinfo  {journal} {Phys. Rev. Lett.}\
  }\textbf {\bibinfo {volume} {94}},\ \bibinfo {pages} {121302} (\bibinfo
  {year} {2005})},\ \Eprint {http://arxiv.org/abs/gr-qc/0408032}
  {arXiv:gr-qc/0408032 [gr-qc]} \BibitemShut {NoStop}%
\bibitem [{\citenamefont {Clarkson}\ and\ \citenamefont
  {Seahra}(2007)}]{Clarkson:2006pq}%
  \BibitemOpen
  \bibfield  {author} {\bibinfo {author} {\bibfnamefont {C.}~\bibnamefont
  {Clarkson}}\ and\ \bibinfo {author} {\bibfnamefont {S.~S.}\ \bibnamefont
  {Seahra}},\ }\href@noop {} {\bibfield  {journal} {\bibinfo  {journal} {Class.
  Quant. Grav.}\ }\textbf {\bibinfo {volume} {24}},\ \bibinfo {pages} {F33}
  (\bibinfo {year} {2007})},\ \Eprint {http://arxiv.org/abs/astro-ph/0610470}
  {arXiv:astro-ph/0610470 [astro-ph]} \BibitemShut {NoStop}%
\bibitem [{\citenamefont {Ishihara}\ and\ \citenamefont
  {Soda}(2007)}]{Ishihara:2007ni}%
  \BibitemOpen
  \bibfield  {author} {\bibinfo {author} {\bibfnamefont {H.}~\bibnamefont
  {Ishihara}}\ and\ \bibinfo {author} {\bibfnamefont {J.}~\bibnamefont
  {Soda}},\ }\href {\doibase 10.1103/PhysRevD.76.064022} {\bibfield  {journal}
  {\bibinfo  {journal} {Phys. Rev.}\ }\textbf {\bibinfo {volume} {D76}},\
  \bibinfo {pages} {064022} (\bibinfo {year} {2007})},\ \Eprint
  {http://arxiv.org/abs/hep-th/0702180} {arXiv:hep-th/0702180 [HEP-TH]}
  \BibitemShut {NoStop}%
\bibitem [{\citenamefont {Crescini}\ \emph {et~al.}(2018)\citenamefont
  {Crescini} \emph {et~al.}}]{Crescini:2018qrz}%
  \BibitemOpen
  \bibfield  {author} {\bibinfo {author} {\bibfnamefont {N.}~\bibnamefont
  {Crescini}} \emph {et~al.},\ }\href@noop {} {\bibfield  {journal} {\bibinfo
  {journal} {Eur. Phys. J.}\ }\textbf {\bibinfo {volume} {C78}},\ \bibinfo
  {pages} {703} (\bibinfo {year} {2018})},\ \bibinfo {note} {[Erratum: Eur.
  Phys. J.C78,no.9,813(2018)]},\ \Eprint {http://arxiv.org/abs/1806.00310}
  {arXiv:1806.00310 [hep-ex]} \BibitemShut {NoStop}%
\bibitem [{\citenamefont {Flower}\ \emph {et~al.}(2018)\citenamefont {Flower},
  \citenamefont {Bourhill}, \citenamefont {Goryachev},\ and\ \citenamefont
  {Tobar}}]{Flower:2018qgb}%
  \BibitemOpen
  \bibfield  {author} {\bibinfo {author} {\bibfnamefont {G.}~\bibnamefont
  {Flower}}, \bibinfo {author} {\bibfnamefont {J.}~\bibnamefont {Bourhill}},
  \bibinfo {author} {\bibfnamefont {M.}~\bibnamefont {Goryachev}}, \ and\
  \bibinfo {author} {\bibfnamefont {M.~E.}\ \bibnamefont {Tobar}},\ }\href@noop
  {} {\  (\bibinfo {year} {2018})},\ \Eprint {http://arxiv.org/abs/1811.09348}
  {arXiv:1811.09348 [physics.ins-det]} \BibitemShut {NoStop}%
\bibitem [{\citenamefont {Manasse}\ and\ \citenamefont
  {Misner}(1963)}]{Manasse:1963zz}%
  \BibitemOpen
  \bibfield  {author} {\bibinfo {author} {\bibfnamefont {F.~K.}\ \bibnamefont
  {Manasse}}\ and\ \bibinfo {author} {\bibfnamefont {C.~W.}\ \bibnamefont
  {Misner}},\ }\href {\doibase 10.1063/1.1724316} {\bibfield  {journal}
  {\bibinfo  {journal} {J. Math. Phys.}\ }\textbf {\bibinfo {volume} {4}},\
  \bibinfo {pages} {735} (\bibinfo {year} {1963})}\BibitemShut {NoStop}%
\bibitem [{\citenamefont {Quach}(2016)}]{Quach:2016uxd}%
  \BibitemOpen
  \bibfield  {author} {\bibinfo {author} {\bibfnamefont {J.~Q.}\ \bibnamefont
  {Quach}},\ }\href@noop {} {\bibfield  {journal} {\bibinfo  {journal} {Phys.
  Rev.}\ }\textbf {\bibinfo {volume} {D93}},\ \bibinfo {pages} {104048}
  (\bibinfo {year} {2016})},\ \Eprint {http://arxiv.org/abs/1605.08316}
  {arXiv:1605.08316 [gr-qc]} \BibitemShut {NoStop}%
\bibitem [{\citenamefont {Holstein}\ and\ \citenamefont
  {Primakoff}(1940)}]{Holstein:1940zp}%
  \BibitemOpen
  \bibfield  {author} {\bibinfo {author} {\bibfnamefont {T.}~\bibnamefont
  {Holstein}}\ and\ \bibinfo {author} {\bibfnamefont {H.}~\bibnamefont
  {Primakoff}},\ }\href@noop {} {\bibfield  {journal} {\bibinfo  {journal}
  {Phys. Rev.}\ }\textbf {\bibinfo {volume} {58}},\ \bibinfo {pages} {1098}
  (\bibinfo {year} {1940})}\BibitemShut {NoStop}%
\bibitem [{\citenamefont {Barbieri}\ \emph {et~al.}(1989)\citenamefont
  {Barbieri}, \citenamefont {Cerdonio}, \citenamefont {Fiorentini},\ and\
  \citenamefont {Vitale}}]{Barbieri:1985cp}%
  \BibitemOpen
  \bibfield  {author} {\bibinfo {author} {\bibfnamefont {R.}~\bibnamefont
  {Barbieri}}, \bibinfo {author} {\bibfnamefont {M.}~\bibnamefont {Cerdonio}},
  \bibinfo {author} {\bibfnamefont {G.}~\bibnamefont {Fiorentini}}, \ and\
  \bibinfo {author} {\bibfnamefont {S.}~\bibnamefont {Vitale}},\ }\href@noop {}
  {\bibfield  {journal} {\bibinfo  {journal} {Phys. Lett.}\ }\textbf {\bibinfo
  {volume} {B226}},\ \bibinfo {pages} {357} (\bibinfo {year}
  {1989})}\BibitemShut {NoStop}%
\bibitem [{\citenamefont {Cruise}\ and\ \citenamefont
  {Ingley}(2006)}]{0264-9381-23-22-007}%
  \BibitemOpen
  \bibfield  {author} {\bibinfo {author} {\bibfnamefont {A.~M.}\ \bibnamefont
  {Cruise}}\ and\ \bibinfo {author} {\bibfnamefont {R.~M.~J.}\ \bibnamefont
  {Ingley}},\ }\href@noop {} {\bibfield  {journal} {\bibinfo  {journal}
  {Classical and Quantum Gravity}\ }\textbf {\bibinfo {volume} {23}},\ \bibinfo
  {pages} {6185} (\bibinfo {year} {2006})}\BibitemShut {NoStop}%
\bibitem [{\citenamefont {Maggiore}(2007)}]{Maggiore:1900zz}%
  \BibitemOpen
  \bibfield  {author} {\bibinfo {author} {\bibfnamefont {M.}~\bibnamefont
  {Maggiore}},\ }\href {http://www.oup.com/uk/catalogue/?ci=9780198570745}
  {\emph {\bibinfo {title} {{Gravitational Waves. Vol. 1: Theory and
  Experiments}}}},\ Oxford Master Series in Physics\ (\bibinfo  {publisher}
  {Oxford University Press},\ \bibinfo {year} {2007})\BibitemShut {NoStop}%
\bibitem [{\citenamefont {Astone}\ \emph {et~al.}(2010)\citenamefont {Astone}
  \emph {et~al.}}]{Astone:2010mr}%
  \BibitemOpen
  \bibfield  {author} {\bibinfo {author} {\bibfnamefont {P.}~\bibnamefont
  {Astone}} \emph {et~al.},\ }\href {\doibase 10.1103/PhysRevD.82.022003}
  {\bibfield  {journal} {\bibinfo  {journal} {Phys. Rev.}\ }\textbf {\bibinfo
  {volume} {D82}},\ \bibinfo {pages} {022003} (\bibinfo {year} {2010})},\
  \Eprint {http://arxiv.org/abs/1002.3515} {arXiv:1002.3515 [gr-qc]}
  \BibitemShut {NoStop}%
\bibitem [{\citenamefont {Tabuchi}\ \emph {et~al.}(2015)\citenamefont {Tabuchi}
  \emph {et~al.}}]{Tabuchi405}%
  \BibitemOpen
  \bibfield  {author} {\bibinfo {author} {\bibfnamefont {Y.}~\bibnamefont
  {Tabuchi}} \emph {et~al.},\ }\href@noop {} {\bibfield  {journal} {\bibinfo
  {journal} {Science}\ }\textbf {\bibinfo {volume} {349}},\ \bibinfo {pages}
  {405} (\bibinfo {year} {2015})}\BibitemShut {NoStop}%
\bibitem [{\citenamefont {Tabuchi}\ \emph {et~al.}(2016)\citenamefont {Tabuchi}
  \emph {et~al.}}]{TABUCHI2016729}%
  \BibitemOpen
  \bibfield  {author} {\bibinfo {author} {\bibfnamefont {Y.}~\bibnamefont
  {Tabuchi}} \emph {et~al.},\ }\href@noop {} {\bibfield  {journal} {\bibinfo
  {journal} {Comptes Rendus Physique}\ }\textbf {\bibinfo {volume} {17}},\
  \bibinfo {pages} {729 } (\bibinfo {year} {2016})},\ \bibinfo {note} {quantum
  microwaves / Micro-ondes quantiques}\BibitemShut {NoStop}%
\bibitem [{\citenamefont {Lachance-Quirion}\ \emph {et~al.}(2017)\citenamefont
  {Lachance-Quirion} \emph {et~al.}}]{Lachance-Quirione1603150}%
  \BibitemOpen
  \bibfield  {author} {\bibinfo {author} {\bibfnamefont {D.}~\bibnamefont
  {Lachance-Quirion}} \emph {et~al.},\ }\href@noop {} {\bibfield  {journal}
  {\bibinfo  {journal} {Science Advances}\ }\textbf {\bibinfo {volume} {3}}
  (\bibinfo {year} {2017})}\BibitemShut {NoStop}%
\end{thebibliography}%

\end{document}